\documentclass[traditabstract]{aa}
\usepackage{graphicx}
\usepackage{txfonts}
\usepackage{amssymb}
\usepackage{lscape}
\DeclareGraphicsExtensions{.pdf,.png,.jpg}
\usepackage{color}

\begin{document}

\def\mpc{h_{75}^{-1} {\rm{Mpc}}} 
\def\kpc{h_{75}^{-1} {\rm{kpc}}}
\newcommand{\mincir}{\raise
-2.truept\hbox{\rlap{\hbox{$\sim$}}\raise5.truept\hbox{$<$}\ }}
\newcommand{\magcir}{\raise
-2.truept\hbox{\rlap{\hbox{$\sim$}}\raise5.truept\hbox{$>$}\ }}
\newcommand{\ha}{\mathrm{H}\alpha}
\newcommand{\hb}{\mathrm{H}\beta}

\title{The XMM spectral catalog of SDSS optically selected Seyfert 2 galaxies}

\author{E. Koulouridis\inst{1,2},
  I. Georgantopoulos\inst{1}, G. Loukaidou\inst{3}, A. Corral\inst{1}, A. Akylas\inst{1}, L. Koutoulidis\inst{1}, E. F. Jim\'enez-Andrade\inst{4,5},
  J.  Le\'on Tavares\inst{5,6}, P. Ranalli\inst{1}}

\institute{Institute for Astronomy \& Astrophysics, Space Applications \&
  Remote Sensing, 
National Observatory of Athens, Palaia Penteli 15236, Greece.
\and Service d'Astrophysique AIM, CEA Saclay, F-91191 Gif sur Yvette, France
\and Faculty of Physics, School of Sciences, Univ. of Athens, Panepistimiopolis, 15771 Ilissia, Greece
\and Argelander-Institut f\"ur Astronomie, Universit\"at Bonn, Auf dem H\"ugel 71, D-53121 Bonn, Germany
\and Instituto Nacional de Astrof\'{\i}sica \'Optica y Electr\'onica (INAOE), Apartado Postal 51 y 216, 72000 Puebla, Mexico
\and Sterrenkundig Observatorium, Universiteit Gent, Krijgslaan 281-S9, B-9000 Gent, Belgium} 
\date{\today}

\abstract{We present an X-ray spectroscopic study of optically selected (SDSS) Seyfert 2 (Sy2) galaxies.
The goal is to study the obscuration of Sy2 galaxies beyond the local universe,
  using good quality X-ray spectra in combination with high S/N optical spectra for their robust classification.
 We analyzed all available XMM-Newton archival observations of narrow emission line galaxies that 
meet the above criteria in the redshift range $0.05<z<0.35$.
 We initially selected narrow line AGN using the SDSS optical spectra and the BPT classification diagram. 
We further modeled and removed the stellar continuum, and we analyzed the residual emission line spectrum
to exclude any possible intermediate-type Seyferts.
 Our final catalog comprises 31 Sy2 galaxies with median redshift $z\sim0.1$. 
X-ray spectroscopy is performed using the available X-ray spectra from the 3XMM and the XMMFITCAT catalogs. 
  Implementing various indicators of obscuration, we find seven ($\sim$23\%) Compton-thick AGN. 
   The X-ray spectroscopic Compton-thick classification agrees with other 
  commonly used diagnostics, such as the X-ray to mid-IR luminosity ratio and the X-ray to [OIII]$\lambda$5007 
  luminosity ratio.  Most importantly, we find four ($\sim$13\%) unobscured Sy2 galaxies, at odds with the simplest unification model. Their
  accretion rates are significantly lower than the rest of our Sy2 sample, in agreement with previous studies that predict
  the absence of the broad line region below a certain Eddington ratio threshold.}

\authorrunning{E. Koulouridis et al.}
\titlerunning{The XMM spectral catalog of SDSS optically selected Seyfert 2 galaxies}
\maketitle

\section{Introduction}

Nearly thirty years ago, the first discovery by Miller \& Antonucci (1983) of
broad permitted emission lines and a clearly non-stellar continuum in the
polarized spectrum of the archetypal Seyfert 2 (Sy2), NGC 1068, was just the beginning
of numerous similar observations in a wide variety of galaxies. Ten years
later, the unification model of active galactic nuclei (AGN)  was formulated upon these observations
(Antonucci 1993). According to the unification model, all AGN are intrinsically identical, while
the only cause of their different observational features is the 
orientation of an obscuring torus with respect to our line of
sight. In more detail, the AGN type depends on the obscuration of 
the broad line region (BLR), a small area at close proximity to the SMBH where the 
broad permitted lines are produced. If the torus happens to be between the observer and the BLR, the optical emission and even the soft X-rays 
are absorbed. Optical spectropolarimetric observations can reveal the hidden broad line region (HBLR) by 
highlighting its scattered emission. 
The observed narrow permitted emission lines are produced at far larger distances 
from the core, where the torus is irrelevant. 
As a prediction of this model, the presence and strength of the broad 
optical emission lines, hence the derived optical spectral type (from type 1 AGN/Sy1s to type 2 AGN/Sy2s, 
and the intermediate types), should correlate with the amount of intervening material as measured in X-rays.

X-ray observations can reveal the exact density of the obscuring torus, even for mildly obscured sources.  
X-ray surveys with {\it Ginga} (Smith and Done 1996) and {\it ASCA} (Turner et al. 1997) 
measured column densities between $10^{22}$ and up to a few times $10^{24}$ $\rm cm^{-2}$ 
in type 2 AGN samples. More recently, Akylas \& Georgantopoulos (2009)
and Brightman \& Nandra (2011), using {\it XMM-Newton},
and Jia et al. (2013, JJ13 hereafter), using {\it Chandra,} also studied 
the obscuration of type 2 X-ray sources
in detail (see also Brandt \& Alexander (2015) for a recent review). 
However, even in the hard X-ray band, the X-ray surveys may be
missing a fraction of highly obscured sources. These sources are called
Compton-thick AGN (see reviews by Comastri 2004 and Georgantopoulos 2013), and they present 
very high obscuring column densities ($>10^{24} cm^{-2}$ , corresponding
to an optical reddening of $A_V>$100).  Even though Compton-thick 
AGN are abundant in the optically selected samples of nearby Seyferts (e.g., Risaliti et al. 1999), 
only a few tens of Compton-thick sources have been identified from X-ray data. Moreover, Krumpe et al. (2008)
found no Compton-thick QSO in their high redshift ($z>0.5$), X-ray selected sample, implying a possible redshift evolution, though 
this may be due to selection.

Although the population of Compton-thick
sources remains elusive, there is concrete evidence of its presence. 
The X-ray background synthesis models can explain the
peak of the X-ray background at 30-40 keV, where most of its
energy density lies, (Frontera et al. 2007; Churazov et al. 2007)
only by invoking a large number of Compton-thick AGN (Gilli
et al. 2007).  We note, however, that other models (e.g., Treister, Urry \& Virani 2009; Akylas et al. 2012) 
succeed in explaining the X-ray background (XRB) spectrum assuming a lower fraction of CT sources.
Additional evidence of a Compton-thick 
population comes from the directly measured space density 
of black holes in the local Universe. It is found that this
space density could be up to a factor of two higher than predicted from
the X-ray luminosity function (Marconi et al. 2004). This immediately 
suggests that the X-ray luminosity function is missing an
appreciable number of obscured AGN.

On the other hand, although widely accepted today, the unification model  cannot explain a series of 
observations. For example, Tran et al. (2001) noticed the absence of a HBLR in polarized light in
many Sy2 galaxies (non-HBLR Sy2 galaxies), suggesting that there is a class of true Sy2 galaxies that intrinsically lack the broad-line 
region (see Ho et al. 2008 for a review). Theoretical models attributed the absence of a BLR to either a 
low Eddington ratio (Nicastro 2000) or to low luminosity (Elitzur \& Shlossman 2006).   Many studies 
propose an evolutionary model where a fraction of Sy2 represents the first or the last phase in the 
life of an AGN (Hunt \& Malkan 1999;
Dultzin-Hacyan, 1999; Krongold et al. 2002; Levenson et al. 2001; Koulouridis et
al. 2006a,b, 2013; Koulouridis 2014, Elitzur, Ho \& Trump 2014). 
This was supported by studies of the local environment of Seyfert galaxies, which showed that 
Sy2s reside in richer environments compared to Sy1s (e.g., Villaroel \& Korn 2014).
 Unobscured low-luminosity Sy2s were detected via investigation 
of their X-ray properties (e.g., Pappa et al. 2000, Panessa \& Bassani 2002; Akylas \& Georgantopoulos 2009).
 Models of galaxy formation also support this scenario: for example, Hopkins et al. (2008) 
assert that the AGN is heavily obscured during its birth. During the build-up of its 
black hole mass, it blows away its cocoon, becoming an unobscured AGN.

In this paper, we compile a sample of bona fide optically selected Sy2 galaxies 
using the SDSS spectra from the data release 10 (DR10). 
   We cross-correlate our sample with the 3XMM/XMMFITCAT spectral catalog (Corral et al. 2015), which contains 
   good quality spectra (at least 50 net counts per XMM detector).
   We identify a sample of 31 Sy2 galaxies with available X-ray spectra 
   in the redshift range z=0.05-0.3.  
   Our study is complemented by X-ray, mid-IR, and [OIII] luminosity ratio diagnostics
    (Georgantopoulos et al. 2013, Trouille \& Barger 2010). 
   This study provides an extension of 
   previous X-ray studies in the local Universe (e.g., Akylas \& Georgantopoulos 2009)
   but also of similar studies at higher redshifts (e.g., JJ13) because of the high S/N X-ray spectra used. 
 
We describe our sample selection in \S2, the X-ray analysis in \$3, while our results
and conclusions are presented in \S4 and \S5, respectively. Throughout
this paper we use $H_0=72$ km/s/Mpc, $\Omega_m=0.27$, and
$\Omega_{\Lambda}=0.73$.

\section{Sample selection}

Our sample is composed of Seyfert 2 galaxies with available X-ray spectra within the XMM-Newton Serendipitous Source catalog 
(Watson et al. 2009, Rosen et al. 2015) and 
optical spectra within the SDSS-DR10.
The names of the sources are taken from the SDSS database. Also, a 
sequence number is given to each source in the current paper (see Table 1). In the diagrams, interesting sources are followed by their sequence numbers.
In the text, the names are followed by the sequence number in parenthesis
to make it easier for the reader to trace the sources in the tables and the diagrams.

\begin{table*}
\begin{minipage}{155mm}
\caption{X-ray observations} 
\tabcolsep 3 pt
\begin{tabular}{llccccccc}
N  &  name  &  obsid  &    ra  &  dec  &  z  & $\rm N_H (\times10^{22}$) & exposure time&  counts\\          
{\em (1)}&{\em (2)}&{\em (3)}&{\em (4)}&{\em (5)}&{\em (6)}&{\em (7)}&{\em
(8)}&{\em (9)}\\
\hline
1  &    J080429.14+235444.1  &  504102101  &  121.1219  &  23.9127  &  0.07432  &  3.18  &    18300/--/--    &    86/0/0                               \\ 
2  &    J080535.00+240950.3  &  203280201  &  121.3961  &  24.1645  &  0.05971  &  3.08  &    5598/8425/--    &    125/83/0                            \\   
3  &    J083139.08+524205.6   &  92800201  &  127.9131  &  52.7016  &  0.05855  &  1.16  &    60280/70920/71790    &    334/127/166                \\
4  &    J084002.36+294902.6  &  504120101  &  130.0095  &  29.8175  &  0.06481  &  1.83  &    17870/22630/22640   &   1072/424/428            \\    
5  &    J085331.05+175339.0    &  305480301  &  133.3791  &  17.8942  &  0.18659  &  2.75  &    34560/--/--    &    173/0/0                           \\      
6  &    J091636.53+301749.3   &  150620301  &  139.1524  &  30.2969  &  0.12339  &  1.25  &    9049/9392/--    &    431/151/0                         \\  
7  &    J100129.41+013633.8  &  302351001  &  150.3724  &  1.6095  &  0.10423  &  3.07  &    31650/42310/42540    &    310/98/127                 \\ 
8 &    J101830.79+000504.9  &  402781401  &  154.6286  &  0.0845  &  0.06233  &  3.00  &    15700/20540/20600    &    753/398/397                  \\  
9 &    J103408.58+600152.1    &  306050701  &  158.5360  &  60.0307  &  0.05101  &  1.51  &    8311/--/11420    &    465/0/133                         \\    
10  &    J103456.37+393941.0  &  506440101  &  158.7349  &  39.6614  &  0.15081  &  1.96  &    68400/83070/83850    &    422/147/120              \\      
11  &    J103515.64+393909.5  &  506440101  &  158.8154  &  39.6527  &  0.10710  &  2.03  &    --/83170/83870    &    0/112/ 92                          \\  
12  &    J104426.70+063753.8  &  405240901  &  161.1109  &  6.6317  &  0.20991  &  3.02  &    24960/--/--    &    92/0/0                               \\   
13  &    J112026.64+431518.4  &  107860201  &  170.1109  &  43.2554  &  0.14591  &  1.32  &    13870/--/--    &    182/0/0                             \\ 
14 &    J113549.08+565708.2   &  504101001  &  173.9555  &  56.9522  &  0.05112  &  1.07  &    17490/21310/21320    &    448/130/127              \\          
15  &    J114826.24+530417.1    &  204260101  &  177.1089  &  53.0717  &  0.09826  &  1.23  &    1701/3632/--    &    112/94/0                         \\  
16  &    J121839.40+470627.6  &  400560301  &  184.6649  &  47.1077  &  0.09390  &  1.00  &    --/37830/37570    &    0/86/137                  \\       
17  &    J123056.11+155212.2  &  112552101  &  187.4978  &  13.5183  &  0.09816  &  2.31  &    8394/--/--    &    80/0/0                                     \\ 
18  &    J122959.45+133105.7    &  106061001  &  187.7338  &  15.87  &  0.18768  &  2.00  &    4660/--/8979    &    96/0/98                                  \\  
19  &    J124214.47+141147.0   &  504240101  &  190.5607  &  14.196  &  0.15710  &  2.22  &    59590/--/80240    &    1465/0/665                       \\  
20 &    J125743.06+273628.2   &  124710201  &  194.4296  &  27.608  &  0.06839  &  1.52  &    30010/--/--    &    119/0/0                                \\   
21  &    J130920.52+212642.7  &  163560101  &  197.3359  &  21.4453  &  0.27858  &  1.57  &    --/28390/28700    &    0/191/215                       \\  
22  &    J131104.66+272807.2   &  21740201  &  197.7694  &  27.469  &  0.23975  &  2.06  &    35000/43000/43100    &    267/73/75                     \\   
23  &    J132525.63+073607.5  &  200730201  &  201.3567  &  7.6022  &  0.12402  &  5.02  &    26900/--/--    &    174/0/0                                \\ 
24  &    J134245.85+403913.6  &  70340701  &  205.6908  &  40.6537  &  0.08926  &  1.57  &    26010/35340/35100    &    393/220/216                \\  
25  &    J135436.29+051524.5  &  404240101  &  208.6515  &  5.2564  &  0.08152  &  7.65  &    11020/--/15780    &    168/0/101                        \\  
26  &    J141602.13+360923.2   &  14862010  &  214.0089  &  36.1567  &  0.17100  &  2.56  &    10910/15780/16140    &    271/146/145                   \\
27  &    J145720.44--011103.6  &  502780601  &  224.3353  &    --1.1844    &  0.08735  &  11.4  &    7942/--/--    &    71/0/0                               \\ 
28  &    J150719.93+002905.0   &  305750801  &  226.8330  &  0.4847  &  0.18219  &  10.5  &    9931/--/--    &    227/0/0                             \\      
29  &    J150754.38+010816.8  &  402781001  &  226.9764  &  1.1381  &  0.06099  &  9.81  &    14330/17900/17890    &    222/81/81              \\      
30  &    J215649.51--074532.4   &  654440101  &  329.2059  &    --7.7589    &  0.05541  &  5.22  &    42310/73600/75600    &    134/63/56             \\
31  &    J224323.18--093105.8  &  503490201  &  340.8464  &    --9.5185    &  0.14509  &  2.72  &    --/113700/114500    &    0/246/247                    \\ 
\hline
\end{tabular}
\tablefoot{{\em (1)} sequence number {\em (2)}, SDSS name,
{\em (3)} XMM-Newton observation ID number, 
{\em (4)} spectroscopic redshift, {\em (5)} right ascension, {\em(6)} declination, 
{\em (7)} Galactic column density in atoms/cm$^2$, {\em (8)} exposure time for the PN/MOS1/MOS2 detectors in seconds, 
{\em (9)} counts on the PN/MOS1/MOS2 detectors.}
\end{minipage}
\end{table*}

\subsection{X-ray selection}

The XMM-Newton catalog is the largest catalog of X-ray sources ever built. Its current 
version, 3XMM-DR4 (http://xmmssc-www.star.le.ac.uk/Catalogue/3XMM-DR4/), contains photometric information for half a million source detections, and
in addition, spectral and timing data for $\sim$ 120000 of them. The count limit adopted by the 3XMM-DR4 pipeline to derive spectral products 
is of 100 EPIC net (background subtracted) counts, in order to allow reliable X-ray spectral extraction and analysis.

 The starting sample was extracted from the XMM-Newton/SDSS-DR7 cross-correlation presented in 
Georgakakis \& Nandra (2011), including more 
than 40000 X-ray sources. We first selected the sources detected in the X-ray hard band (2-8 keV), a band less 
affected by obscuration than is the soft one (0.5-2 keV). A total of 1275 sources were found to have available optical spectra within SDSS-DR7. Out of these,
1018 sources had available 3XMM-DR4 spectral data. The corresponding SDSS optical spectra of
these 1018 sources were manually examined in order to identify Seyfert 2 galaxies, resulting in our final sample of Sy2s (see next section). 
It is worth noting that two of these sources have more than one XMM-Newton observation with spectra within 3XMM-DR4, from which we used the longest one.

\begin{table*}
\begin{minipage}{175mm}
\caption{X-ray spectral analysis} 
\tabcolsep 3 pt
\centering
\def\arraystretch{1.3}
\begin{tabular}{lcccccccccccc} \\ \hline
N  &  $\rm N_H $  & $\Gamma_{soft}$  & $\Gamma_{hard}$   &  EW   &  flux  &  $L_X$ &  p1/p2 &  cstat/dof & $L_{[OIII]}$&$L_{12}$ & $L_{bol}$&log($M_{BH}$)\\
  &  $(\times10^{22}$)   &  &    &    &  $(\times10^{-14})$  &  $(\times10^{43})$ &  &  & $(\times10^{42})$&$(\times10^{43})$&$(\times10^{43})$&\\
{\em (1)}&{\em (2)}&{\em (3)}&{\em (4)}&{\em (5)}&{\em (6)}&{\em (7)}&{\em(8)}&{\em(9)}&{\em(10)}&{\em(11)}&{\em(12)}&{\em(13)}\\
\hline
1 &       $18.84_{-5.56}^{+7.68}$       &      $ 7.6_{-4.6}^{+7.6}$       &     1.8$\ddagger$     &$  <0.33      $                        &22.2&0.27&0.001&       105.2/96          &0.18       &   0.4   &  1.1 &  6.9        \\   
2 &       $45.83_{-13.94}^{+21.30}$       &     $  2.4_{-0.4}^{+0.4}$       &  1.8$\ddagger$    & $  0.29_{-0.24}^{+0.71}$       &25.0&0.20&0.015&       165.25/210    &0.60                         &    0.2    & 0.3 & 6.8      \\     
3 &       $23.56_{-3.79}^{+4.00}  $     &       $2.1_{-0.4}^{+0.4}$       &  $2.1_{-0.4}^{+0.4}$      & $  0.26_{-0.13}^{+0.20}$       &9.3&0.07&0.012&       537.63/618     &0.17                        &    0.7    &   1.0 & 6.3  \\    
4 &       $54.24_{-5.82}^{+5.32} $      &       $2.8_{-0.1}^{+0.2}$       &   $ 1.4_{-0.7}^{+0.7}$      & $  0.30_{-0.80}^{+0.80}$       &69.7&0.63&0.024&       1088.68/1403    &1.11  &    8.8 & 19.9 & 7.6 \\      
5 &       $26.38_{-20.54}^{+27.12} $      &      $ 2.5_{-0.7}^{+0.7}$ &  1.8$\ddagger$   &$  <0.17 $&17.8&1.20&0.293&       147.19/189              &5.77                                            &   11.4    & 25.9 & 7.9      \\        
6 &       $<0.06         $      &      \multicolumn{2}{c}{$ 1.8_{-0.1}^{+0.2}$}   & $ <0.74  $                          &22.4&0.87&                &       425.68/522               &3.47       &   1.8    &      2.8    &    8.2  \\    
7 &       $4.75_{-1.40}^{+1.28} $      &      $ 1.4_{-0.4}^{+0.4}$ &  $1.4_{-0.4}^{+0.4}$    &        $   0.29_{-0.17}^{+0.20}$       &15.6&0.39&0.018&       465.96/529      &0.07                      &   1.4    & 1.9 & 7.2      \\   
8&       $2.34_{-0.32}^{+0.38}  $      &     \multicolumn{2}{c}{$1.5_{-0.2}^{+0.2}$}   & $  <0.10      $                        &60.1&0.54&                   &  825.83/1078     &0.04       &    0.3  &  0.4 & 6.2   \\    
9$^\dagger$&       $30.32_{-11.62}^{+19.24} $      &      $ 2.9_{-0.2}^{+0.2} $      &  1.8$\ddagger$   &  $  1.30_{-0.48}^{+0.80}$       &17.8&0.07&0.293&       430.95/474    &4.98                     &    7.4   &  10.3 & 8.2  \\     
10&       $64.58_{-23.33}^{+29.99} $      &      $ 3.0_{-0.2}^{+0.2}$ &    1.8$\ddagger$    &     $  0.52_{-0.25}^{+0.26}$        &4.1&0.20&0.102&       505.46/546    &5.37                    &   4.4   &  6.2 &  8.2   \\        
11&       $14.35_{-5.85}^{+6.61}$       &      $ 6.3_{-1.5}^{+2.1}$&  $ 2.0_{-1.1}^{+1.2}$     &       $ <1.21   $                          &9.6&0.25&0.007&       189.43/210 &0.22   &   7.4    & 16.7 & 7.6        \\    
12&       $90.26_{-48.60}^{+31.61} $      &      $ 1.5_{-0.9}^{+0.9}$       &  $1.5_{-0.9}^{+0.9}$ & $   <13.68  $                         &11.6&0.92&0.004&       82.28/98             &1.90            &   87.9    & 118.0  &  8.7        \\     
13&       $5.49_{-2.51}^{+2.68} $      &      $ 1.3_{-0.7}^{+0.6}$       &   $1.3_{-0.7}^{+0.6}$   &  $   0.32_{-0.28}^{+0.85}$       &19.1&0.95&0.054&       140.57/186       &0.23                     &   2.8   & 6.3 &6.0           \\   
14&       $130.46_{-56.34}^{+69.32}$       &     $  2.9_{-0.2}^{+0.2}$       &  1.8$\ddagger$   & $  <0.65     $                        &7.6&0.04&0.006&       434.45/462           &6.62           &    22.2  & 50.8 & 7.6   \\     
15&       $1.92_{-0.69}^{+0.88}  $  &      \multicolumn{2}{c}{$ 1.8_{-0.5}^{+0.6}$}   &         $  <0.48  $                         &144&3.37&              &       169.96/204              &0.64      &   0.9    & 1.2  & 7.3       \\    
16$^\dagger$&       $16.07_{-10.34}^{+39.22} $      &      $ 2.9_{-0.4}^{+0.4}$ &   1.8$\ddagger$     &      $ 0.85_{-0.66}^{+0.75}$        &7.9&0.16&0.173&       145.04/155    &5.47                    &   9.4 &  13.1  &  7.3  \\         
17&       $4.83_{-2.48}^{+2.80}  $     &      $ 2.3_{-1.1}^{+1.0}$&   $2.3_{-1.1}^{+1.0}$    &$\star$&12.6&0.30&0.015&       82.11/85                   &2.06                                                  &   4.8   &   6.3  & 7.4      \\   
18&       $1.81_{-0.87}^{+1.08} $  &    \multicolumn{2}{c}{$ 1.2_{-0.5}^{+0.5}$}    & $  <0.35  $                         &45.2&3.81&               &       183.1/184                &0.59      &   9.3  & 15.0 & 6.5      \\    
19&       $<0.04                $      &      \multicolumn{2}{c}{$ 1.8_{-0.1}^{+0.1}$}   &  $  0.38_{-0.21}^{+0.21}$        &9.9&0.65&               &       748.87/885       &1.45              &   1.4     &  4.0  &      8.0    \\    
20&       $12.62_{-7.20}^{+6.96} $      &       $1.7_{-1.1}^{+1.0}$       &   $1.7_{-1.1}^{+1.0}$     & $ 0.35_{-0.35}^{+0.51}$        &9.7&0.10&0.052&       140.82/167      &0.04                       &    0.3    & 0.4 & 5.9      \\     
21&       $<0.08     $                 &      \multicolumn{2}{c}{$ 2.2_{-0.2}^{+0.2}$}    &$\star$&4.7&1.21&                &       214.46/247                         &0.10                                   &   0.5    &     0.5    &    6.9   \\    
22$^\dagger$&       $243.07_{-115.26}^{+303.45}$      &     $ 2.6_{-0.2}^{+0.2}$     &  1.8$\ddagger$  &  $ 0.65_{-0.60}^{+0.81}$         &4.7&0.37&0.003&       296.79/365     &3.07                        &   11.7   &  23.3  &   7.8    \\     
23&       $0.39_{-0.22}^{+0.39} $      &      \multicolumn{2}{c}{$ 1.5_{-0.4}^{+0.5}$}       &  $   <0.88$                          &16.5&0.62&              &       172.56/170         &0.13        &   0.4   &   1.1   &     7.4         \\   
24&       $6.47_{-1.07}^{+1.31} $      &      $ 2.0_{-0.3}^{+0.4}$ &  $2.0_{-0.3}^{+0.4}$    &         $  0.17_{-0.13}^{+0.17}$       &41.1&0.77&0.009&       659.27/707       &0.26                     &   2.3    & 3.4 & 7.1     \\    
25$^\dagger$&       $<0.07 $                      &      \multicolumn{2}{c}{$ 0.8_{-0.2}^{+0.2}$}  &$<0.70$&22.7&0.34&&       274.4/258             &0.16                                                      &   0.9    & 2.5 & 6.3           \\    
26&       $1.98_{-0.47}^{+0.50} $   &   \multicolumn{2}{c}{$ 1.7_{-0.3}^{+0.3}$}  &  $  <0.37  $                          &46.5&3.47&               &       383.72/507                &1.48     &   42.3   &  97.8 & 8.1       \\  
27&       $5.29_{-3.53}^{+4.69} $      &      \multicolumn{2}{c}{$ 1.6_{-1.1}^{+1.2}$}  &$<0.60$&13.1&0.23&                   &       97.74/80           &0.22                                       &   1.4       &   1.9 & 7.1             \\   
28&       $28.78_{-12.22}^{+15.82} $      &      $ 1.7_{-0.7}^{+0.5}$      & $1.7_{-0.7}^{+0.5}$    & $  <0.46 $                          &34.6&2.56&0.045&       246.1/310           &10.66             &   2.4   & 32.3 & 8.8    \\       
29$^\dagger$&       $32.18_{-12.68}^{+19.90}$       &     $  3.4_{-0.3}^{+0.4}$       & 1.8$\ddagger$    & $  1.22_{-0.73}^{+1.89}$       &10.6&0.09&0.155&       339.5/354    &1.71                         &    1.5  & 2.1 & 7.4    \\        
30$^\dagger$&       $14.95_{-7.28}^{+11.92}  $     &       $3.6_{-0.4}^{+0.5} $       &   1.8$\ddagger$     &  $  2.08_{-1.24}^{+3.05}$       &3.8&0.03&0.280&       266.06/300    &0.96                   &    2.7   & 4.9 & 7.4      \\         
31&       $2.61_{-0.68}^{+0.89} $    &     \multicolumn{2}{c}{ $ 1.3_{-0.3}^{+0.4}$}   &  $  <0.25  $                          &10.3&0.51&               &       355.49/417           &0.45      &   0.2   &  0.4 &   6.2         \\   
\hline
\end{tabular}
\tablefoot{{\em (1)} sequence number, {\em (2)}, obscuring column density in atoms/cm$^2$,
{\em (3)-(4)} power-law photon index of the scattered and the continuum emission, respectively 
(in the case of a single power-law fit. the value is listed in the middle of the two columns),
{\em (5)} equivalent width in keV of the FeK$\alpha$ line at 6.4 keV, {\em (6)} X-ray flux in $\rm erg\; s^{-1} cm^{-2}$, {\em
(7)} X-ray luminosity in $\rm erg\; s^{-1}$, 
{\em (8)} the ratio of the scattered to the continuum emission normalization, 
{\em (9)} C-statistics and degrees of freedom, {\em (10)} reddening-corrected [OIII] luminosity in $\rm erg\; s^{-1}$,
{\em (12)} AGN bolometric luminosity computed from the SED, {\em (13)} 
black hole mass computed from the $M_{BH}-\sigma*$ relation. 
\newline$\star$ Not constrained. \newline$\dagger$ See appendix for notes on individual objects. \newline $\ddagger$ Fixed $\Gamma_{hard}$.}

\end{minipage}
\end{table*}

\subsection{Optical selection}

We built the final Sy2 sample based on the emission line properties of their SDSS optical spectra. Initially,
we selected only emission line galaxies with redshifts between $z$=0.05 and $z$=0.35. The lower redshift limit excludes 
all already extensively studied and well-known Seyferts (e.g., Akylas \& Georgantopoulos 2009), while the upper limit 
ensures that the $\ha$ and [NII] emission lines are within the SDSS spectral range.
Furthermore, we excluded all objects where the velocity dispersion of the $\rm H\alpha$ line is greater than 500 km/s,
since these objects are certainly broadline AGN. The rest of the objects were placed on a BPT diagram (Baldwin, Phillips, and Terlevich, 1981) and 
star-forming galaxies, composite galaxies, and LINERS were removed according to the criteria of Kewley et al. (2001) and Schawinski et al. (2007).

We used the MPA-JHU emission line fluxes published in DR8 (Brinchmann et al. 2004; Tremonti et al. 2004), although DR10 also contains 
data from the recent spectroscopic analysis of the Portsmouth Group (Thomas et al. 2013). However, the latter includes only 
those galaxies from the first two years of observations of the SDSS-III/Baryonic Oscillation Spectroscopic Survey
(BOSS) collaboration. We note that a comparison between the two databases 
by Thomas et al. (2013) has shown that the discrepancy between the calculated emission line fluxes is small. However, the comparison 
was made after rescaling the Portsmouth values with a factor provided by the ``spectofiber'' keyword in the MPA-JHU database. This 
rescaling was originally applied to the MPA-JHU data so that the synthetic r-band magnitude computed from the spectrum matches 
the r-band fiber magnitude measured by the photometric pipeline.
The use of either database does not significantly affect the BPT diagram, since we only need the emission line ratios. 

 We note that in some cases the broadening of the Balmer lines cannot be automatically detected (Seyfert 1.5, 1.8, and especially 1.9),
since it only affects the lower part of the lines. As a result, the automated modeling of the line by a single Gaussian may result in lower
velocity dispersion values than what is expected from a broad line profile, and the source may be misclassified as a narrow-line AGN.
However, since we sought a broad-line-free sample, 
the spectra of all remaining AGN were eye-inspected with the ``interactive spectrum'' tool of the SDSS, and all evident intermediate-type 
Seyferts were removed. 
After the above filtering, the catalog of Sy2s included 40 objects.

Despite the above selection, a number of sources in our sample still have discrepant classifications in the literature; 
i.e., eight of the sources are listed as Sy1s in Veron-Cetty \& Veron (2010, V\&V10 hereafter) catalog, 
plus another one in the NED (NASA extragalactic database). Although none of these objects can actually be a Sy1, 
we proceeded with our own optical spectrum analysis to determine whether there is any broadening of the permitted emission lines.

\subsection{Optical spectrum analysis}

The spectra have been retrieved from the SDSS-DR10 and corrected for Galactic extinction using the maps of Schlegel (1998).
We use the stellar population synthesis code \textmd{STARLIGHT}\footnote{http://www.starlight.ufsc.br/} to obtain the best fit to an observed 
spectrum $O_{\lambda}$, taking the corresponding flux error into account. The best fit is a combination of single stellar populations (SSP) 
from the evolutionary synthesis models of  (Bruzual 2003) and a set of  power laws to represent the AGN continuum emission. 
Following the latter approach, several studies have been successful at disentangling the host galaxy and AGN emission components in SDSS 
spectra (Cid-Fernandes 2011; Tavares 2011).

We use a base of 150 SSPs plus six power laws in the form F($\lambda$) = 10$^{20}$($\lambda$ / 4020)$^\beta$, 
where $\beta$= -0.5, -1, -1.5, -2, -2.5, -3.  Each SSP spans six metallicities, Z = 0.005, 0.02, 0.2, 0.4, 1, and 2.5, 
$Z_{\odot}$, with 25 different ages between 1~Myr and 18~Gyr. Extinction in the galaxy is taken into account in the synthesis, 
assuming that it arises from a foreground screen with the extinction law of (Cardelli 1989). The code finds the minimum $\chi^{2}$,
\begin{equation}
 \chi^{2} = \sum_{\lambda} \left(  \frac{O_{\lambda}- M_{\lambda}} {\sigma_{obs}}\right)
,\end{equation}
where $M_{\lambda}$ is the model spectrum (SSP and power laws),
obtaining the corresponding physical parameters of the modeled
spectrum: star formation history, $x_{j}$, as a function of a base
of SSP models normalized at $\lambda_{0}$, $b_{j,\lambda}$,
extinction coefficient of predefined extinction laws,
$r_{\lambda}$, and velocity dispersion $\sigma_{\star}$, which obeys
the relation

  \begin{equation}
   M_{\lambda}= M_{\lambda 0} \left ( \sum_{j=1}^{N_{SSP}} x_{j}, b_{j,
   \lambda} r_{\lambda} \right) \otimes G(v_{\star}, \sigma_{\star})
  .\end{equation}

\noindent A detailed description of the \textmd{STARLIGHT} code can be found in the publications of the SEAGal
collaboration (Cid-Fernandes 2005 ; Cid-Fernandes 2007; Mateus 2006, Asari 2007). In Fig. 1 we present two 
examples of the spectral decomposition results.

\begin{figure*}
\resizebox{18cm}{6cm}{\includegraphics{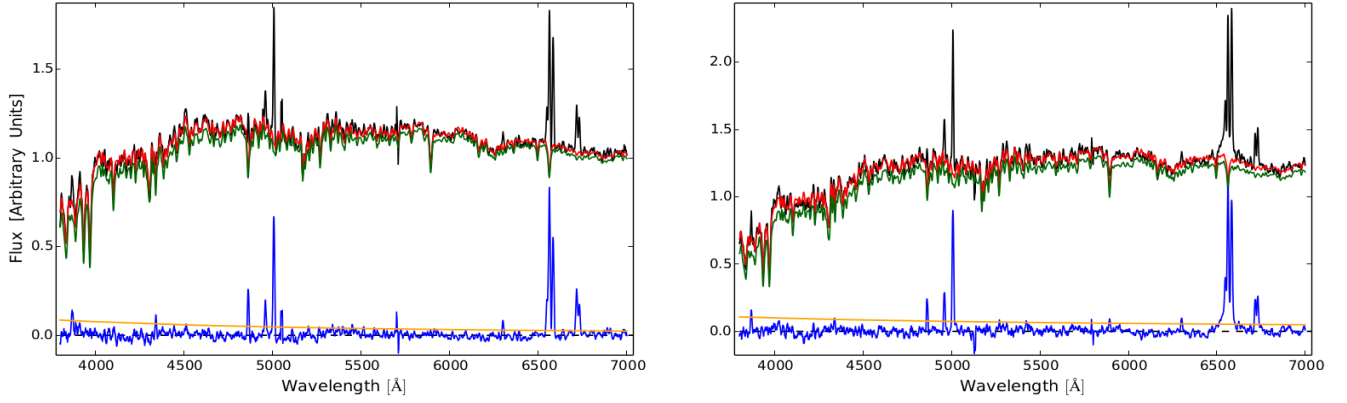}}
\caption{Two examples of the optical spectrum analysis using STARLIGHT software. 
We plot the observed spectrum (black), the host galaxy model spectrum (green), the modeled spectrum (red), 
the AGN continuum emission (yellow), and the residual emission-line-spectrum (blue)}
\end{figure*}

After subtracting the stellar background, we use the commercial software {\tt PEAKFIT}, by {\it Systat Software Inc.}, to model the 
emission lines. We analyze separately the red ($\rm H\alpha$, N[II] and S[II] emission lines) and the blue ($\rm H\beta$ and [OIII] emission lines)
parts of the spectrum. We initially model the emission lines in the blue part, since we are mostly interested in the profile 
of the [OIII]${\lambda5007}$ narrow emission line, with which we also try to fit the lines in the blue part 
and especially the $\rm H\alpha$. We model the [OIII] line with a mixed Gaussian and Lorentzian profile. The contribution of each profile to the fit 
is a free parameter. If the same profile can also be applied to the red part of the spectrum, we consider this source 
as a narrow line AGN and keep it in our sample. If there is still a need for an extra broad component to model the $\rm H\alpha$ the source is
discarded. In any case, the [NII]${\lambda6583}$/[NII]${\lambda6548}$ flux ratio should be $\sim3$. 
We find that seven out of the 40 sources present a broad $\rm H\alpha$ component. Most of these sources belong to the list of ambiguous-type Seyferts that
we described in the previous section.

Finally, we plot the BPT diagram anew, this time with the line ratios calculated by the above spectral analysis. Although the differences are small,
we find that a source that was already close to the AGN-LINER separating line, falls in the LINER region and is therefore excluded. The BPT diagram
is plotted in Fig. 2.

\begin{figure}
\resizebox{9cm}{9cm}{\includegraphics{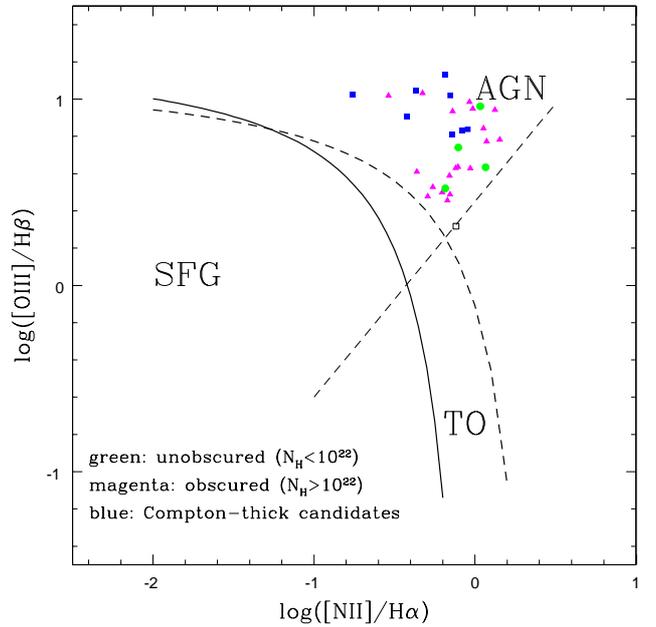}}
\caption{BPT diagnostic diagram for the Sy2 sample. The various levels 
of obscuration are color-coded. The continuous curve  
denotes the star-forming - AGN separation line of Kauffmann et al. (2003), and the thick dashed curve the respective line of Kewley et al. (2001). 
The dashed straight line denotes the LINER - AGN threshold by Schawinski et al. (2007). Composite or 
transition objects (TO) between the AGN and star-forming phase are found in the area between the two curves. The open square indicates the position 
of a discarded LINER.}
\end{figure}

\section{X-ray spectral fitting}

The X-ray data have been obtained with the EPIC (European
Photon Imaging Cameras, Str\"uder et al. 2001; Turner et al. 2001)
onboard XMM-Newton. X-ray photons are collected by
three detectors (PN, MOS1, and MOS2). All available instrument spectra are modeled simultaneously by using XSPEC, 
the standard package for X-ray spectral analysis (Arnaud 1996).
We used Cash statistics (C-statistics), implemented as cstat in XSPEC to obtain reliable spectral-fitting results 
even for the lowest quality spectra in our sample. Many of our sources were detected in only one or two
of the three detectors (see Table 1).

\renewcommand{\baselinestretch}{1.2}
The X-ray spectra of type 2 AGN are usually complicated
and consist of multiple components: power-law, thermal,
scattering, reflection, and emission lines (see Turner et al. 1997;
Risaliti 2002; Ptak et al. 2006; LaMassa et al. 2009). Therefore,
no single model could successfully fit the spectra in all cases. We initially 
tried to model all spectra with a single absorbed power law, but if the 
fit was not acceptable we added a second power law. Since a strong line is expected 
in obscured sources, we fit a Gaussian line for the FeK$\alpha$ emission line in both cases. 
In more detail, this includes:

\begin{itemize}

\item{Single absorbed power law plus Gaussian FeK$\alpha$ line.}\newline

We assumed a standard power-law model with two absorption
components (wabs*zwabs*pow in XSPEC notation) to fit the
source continuum emission. The first component models the Galactic absorption. 
Its fixed values are obtained from
Dickey \& Lockman (1990) and are listed in Table 1. The second component represents the AGN intrinsic absorption
and is left as a free parameter during the modeling procedure. A Gaussian component has also been included to describe
the FeK$\alpha$ emission line. We fix the line energy at 6.4 keV in the source rest frame (except in the case of J090036.85+205340.3 (N6) 
where the line was found at 6.7 keV and implies ionized Fe)
and the line width $\sigma$ at 0.01 keV ($\sim$10\% of the instrumental line resolution
of XMM-Newton). In 12 cases the fitting procedure gives a rejection probability less
than 90 per cent and we can accept the model. However,
when this simple parametrization is not sufficient to model the
whole spectrum, additional components must be included as described in the next paragraph. 
\\
\item {Double power law plus Gaussian FeK$\alpha$ line.}\newline

 In the remaining 20 cases, an additional power law was necessary to obtain an acceptable fit (wabs*(pow+zwabs*pow), in XSPEC notation). 
 The additional power law is only absorbed by the galactic column density.
Initially, the photon indices of the soft (scattered/unabsorbed) and hard (intrinsic/absorbed) power-law components were 
tied together. However, in 13 cases the value of the hard power-law photon index $\Gamma_{hard}$ was too high 
(the average photon index of the intrinsic power-law measured in AGN is usually $\sim1.8-2$), 
and we needed to untie it from the soft one to obtain an acceptable fit. In the cases where the data quality was not high 
enough to constrain $\Gamma_{hard}$, we fixed it to 1.8 (see Table 2). 

\end{itemize}

The X-ray analysis revealed that one of the sources is the brightest galaxy 
of a contaminating X-ray luminous cluster. We chose to exclude this source from our sample since
we cannot provide any reliable X-ray measurements. Our final sample comprises 31 Seyfert 2. In Fig. 3 we present some examples of the X-ray spectra of 
unobscured ($<10^{22}cm^{-2}$, left panels) and strong FeK$\alpha$-line sources (right panels).
 
\begin{figure*}
\centering
\resizebox{18cm}{11cm}{\includegraphics{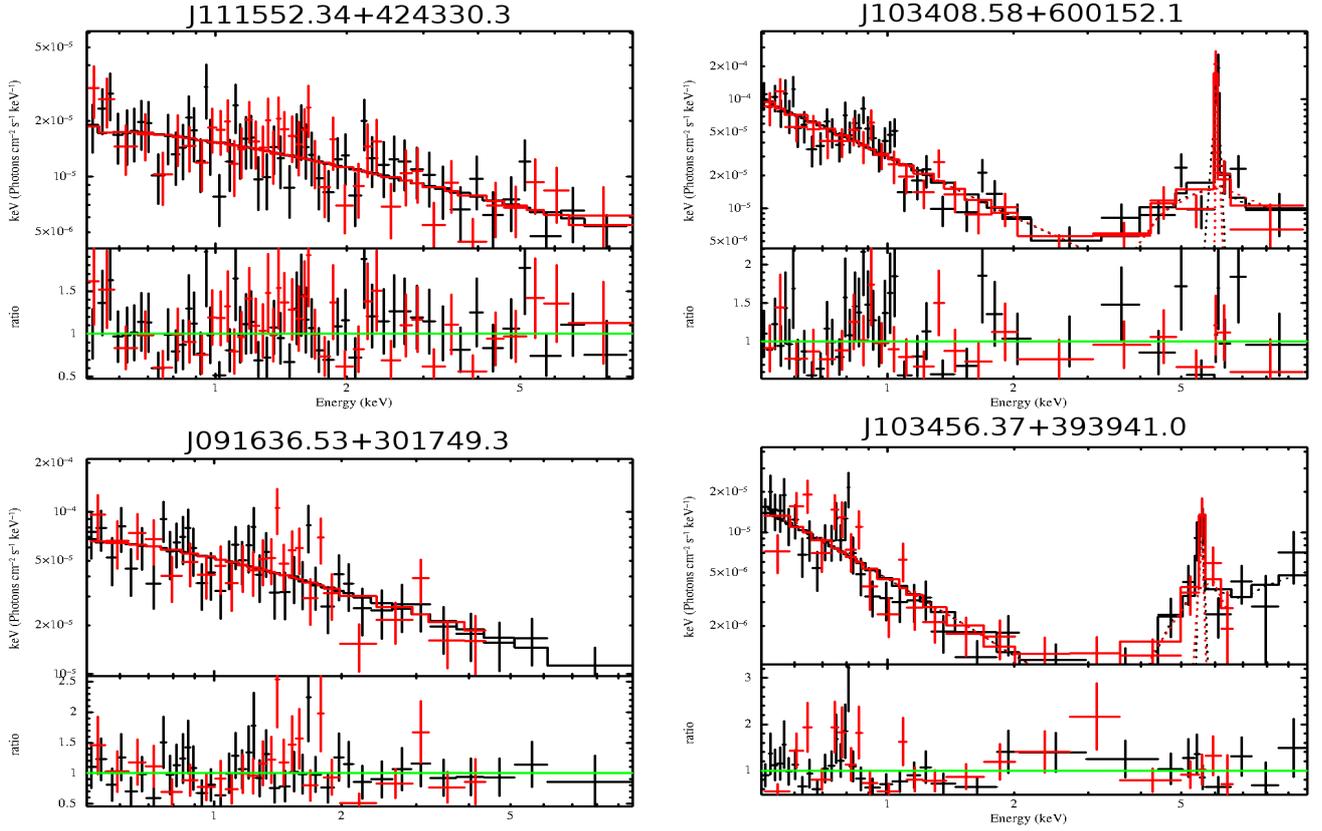}}
\caption{Left panels: X-ray spectral modeling of two unobscured sources with a single power law. Right panels: X-ray spectral modeling of 
two candidate Compton-thick sources with a double power law and an iron FeK$\alpha$ line. 
The black points and the black line denote the PN detector, while the red line denotes the merged MOS1 and MOS2.}
\end{figure*}

\section{Results}

In the next sections we use various criteria and diagnostic diagrams
to investigate the possibility that some objects are more obscured than
we can infer from their $\rm N_H$ values and that Compton-thick candidate sources 
are indeed heavily obscured. 

\subsection{Candidate Compton-thick sources}

Only two of the sources have $\rm N_H>10^{24}$ cm$^{-2}$, consistent with the high values that define Compton-thick sources. 
Also, sources (N10) and (N12) are consistent with being CT within the uncertainties. However, except for the column density as a direct indicator of obscuration, there
are other criteria, based not only on the X-ray but also on the 
optical and the infrared emission, that could point to possible Compton-thick sources 
within our sample. In more detail, a heavily obscured source can have one
or more of the following characteristics:

\begin{enumerate}
\item Flat X-ray spectrum ($\Gamma<1$). This implies the presence of a
strong reflection component that intrinsically flattens the
X-ray spectrum at higher energies (e.g., Matt et al. 2000).
\item High equivalent width of the FeK$\alpha$ line ($\sim$1 keV). In this case 
a Compton-thick nucleus is evident since 
the line is measured against a heavily obscured continuum 
(Leahy \& Creighton 1993) or only against the reflected component.
\item Low X-ray to mid-infrared ($\rm L_{12}$) luminosity ratio.  All
Compton-thick sources should have low $\rm L_{2-10 keV}$ to $\rm L_{12}$ ratios, since 
the mid-IR luminosity of an AGN should be dominated by very hot dust
and the X-ray emission should be suppressed by high amounts of absorption (e.g., Lutz et al. 2004; Maiolino et al. 2007).
\item Low X-ray to optical luminosity ratio. The [OIII] line emission originates in the narrow line
region and is not affected by the circum-nuclear obscuration. 
Therefore, the ratio between the observed hard X-ray (2-10 keV) and
[OIII] line luminosity could be used as an indicator of the
obscuration of the hard X-ray emission (Mulchaey et al. 1994;
Heckman et al. 2005; Panessa et al. 2006; Lamastra et al.
2009; LaMassa et al. 2009; Trouille \& Barger2010).
\end{enumerate}

\subsubsection{Flat X-ray spectrum as an indicator of obscuration}

The first criterion of $\Gamma$ $<1$ is satisfied only by J135436.29+051524.5 (N25). However, this source 
cannot be included in the Compton-thick candidate sources because there is evidence of partial covering.
For more detail see the notes on individual objects in the appendix.

\subsubsection{High equivalent width of the FeK$\alpha$ line as an indicator of obscuration}

The second criterion of a strong FeK$\alpha$ line is satisfied by four objects (see Table 2). Although the presence of the strong line provides robust 
evidence of their obscuration, all four exhibit lower $\rm N_H$ values 
than what is expected by a Compton-thick source. Therefore, we 
also fit these sources with the model of Brightman \& Nandra (2011), which is based on Monte-Carlo simulations. 
The advantage of this model is that it fits an iron line consistently with the computed $\rm N_H$. Thus, it cannot result in a good fit with 
a low $\rm N_H$ value and at the same time a high-EW iron line, and vice versa. 
The fitting confirms that these four sources are indeed Compton-thick. More details can be found in the notes on individual objects in the appendix.
Therefore, we do include them in our list of CT sources.

Also, we need to examine the X-ray spectra of the unobscured sources carefully for the FeK$\alpha$ line 
that could give away the presence of obscuration. However, as we can see in Table 2, the line is actually detected only in one 
out of the five sources, and the equivalent width (EW) is relatively small ($0.38^{+0.21}_{-0.21}$). We do not detect  
the line in the spectra of any other unobscured source, and the given value of the EW is just the upper limit. Thus, there is no evidence of obscuration 
based on the the presence of a FeK$\alpha$ line.

We note that this criterion is not explicit. High equivalent width 
lines may also appear in the case of anisotropic distribution of the scattering medium
(Ghisellini et al. 1991) or in the case of a time
lag between the reprocessed and the direct component (e.g.,
NGC 2992, Weaver et al. 1996). On the other hand, Compton-thick sources 
with FeK$\alpha$ EW well below 1 keV have been reported
(e.g., Awaki et al. 2000, for Mkn1210).

\subsubsection{The $L_X/L_{12}$ ratio as an indicator of obscuration}

The detection of a low X-ray to mid-IR luminosity ratio has
been widely used as the main instrument for detecting
faint Compton-thick AGN, which cannot be easily identified in
X-ray wavelengths (e.g., Goulding et al. 2011). This is because
the mid-IR luminosity (e.g., 12 $\mu $m or 6 $\mu $m) is a good proxy
for the AGN power because it should be dominated by very hot dust
that is heated by the AGN (e.g., Lutz et al. 2004; Maiolino et al. 2007). 
At these wavelengths, the contribution of the stellar light and of colder dust heated by young stars should be small.
Gandhi et al. (2009) presented high angular resolution mid-IR (12 $\mu $m) observations 
of the nuclei of 42 nearby Seyfert galaxies. These observations provide the least contaminated core fluxes of AGN.
These authors find a tight correlation between the near-IR fluxes and the intrinsic X-ray luminosity (the Gandhi relation). 

Spitzer observations do not have the spatial resolution to resolve the core, and 
the infrared luminosity of an AGN is probably contaminated 
by the stellar background and the star-forming activity of the galaxy.
To obtain an estimate of the purely nuclear 12 $\mu $m infrared luminosity of our sources,
we constructed their spectral energy distributions (SED) and computed the various contributions. 
To model the spectra we used optical data from the SDSS (five optical bands), photometry in the four
WISE bands (3.4, 4.6, 12, and 22 $\mu $m) (Wright et al. 2010), and photometry in the three 2MASS
bands (J, H, and K) for all sources. Although WISE does include the 12 $\mu $m band, we are only interested 
in the AGN contribution, so that the construction of the SED and the decomposition of the AGN and host galaxy 
component is essential. For more details about the code used, the interested reader should refer 
to Rovilos et al. (2014, Appendix A).

In Fig. 4 we present the obscured X-ray luminosities against the 12 $\mu$m luminosities. 
All our unobscured sources seem to follow the Gandhi-relation closely, and none of them shows 
unusually high infrared luminosity compared to the X-ray. On the other hand, 
candidate CT sources are found closer to the dashed line that demarcates the purely CT region. 
The sources located below this line are all candidate CT according to our analysis. 
Therefore, it is is unlikely that we are missing any CT candidates among the Sy2 sample.

\begin{figure}
\resizebox{9cm}{9cm}{\includegraphics{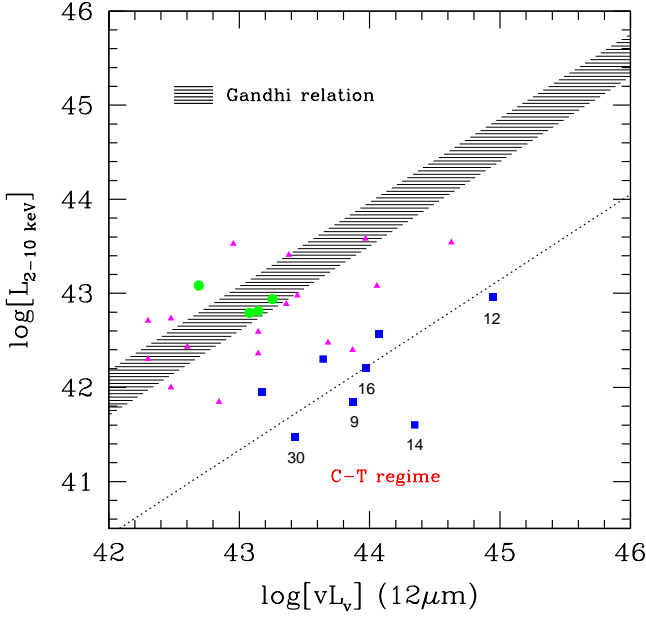}}
\caption{Absorbed X-ray (2-10 keV) X-ray luminosity against the 12$\mu$ m luminosity. Color- and shape-coding 
as in Fig. 1. The typical errors are on the order of 30\% and 20\% for the IR and X-ray luminosity, respectively, 
including the uncertainties in the model fitting. The hatched diagram represents the 1 $\sigma$
envelope of the local (Gandhi et al. 2009) relation. The dotted line corresponds to a factor of 30 lower X-ray 
luminosity as is typical in many Compton-thick nuclei. The numbers of interesting sources 
correspond to the sequence numbers in Tables 1 and 2.}
\end{figure}

\begin{table}
\begin{minipage}{87mm}
\caption{Candidate Compton-thick criteria} 
\tabcolsep 3 pt
\begin{tabular}{lccccc}
N  &  name  & $\rm N_H$ & FeK$\alpha$ & $\rm L_x/L_{12}$ &  $\rm L_X/L_{[OIII]}$\\          
{\em (1)}&{\em (2)}&{\em (3)}&{\em (4)}&{\em (5)}&{\em (6)}\\
\hline   
9 &    J103408.58+600152.1    & $>10^{24}\dagger$ &  x        &  x   &  x   \\    
10  &    J103456.37+393941.0  &   $>5\times10^{23}$ &  &      &     \\      
12  &    J104426.70+063753.8  &   $>9\times10^{23}$ &  &  x   &     \\   
14 &    J113549.08+565708.2   &   $>10^{24}$        &  &  x   &  x   \\          
16  &    J121839.40+470627.6  &  $>10^{24}\dagger$  &  x         &  x   &   x \\       
22  &    J131104.66+272807.2   &  $>10^{24}$   &       &      &  x \\   
29  &    J150754.38+010816.8  & $>10^{24}\dagger$  &  x     &   &     \\      
30  &    J215649.51--074532.4   & $>10^{24}\dagger$ & x    &  x  &  x \\
\hline
\end{tabular}
\tablefoot{{\em (1)} sequence number {\em (2)}, SDSS name,
{\em (3)} column density $\rm N_H$ in $cm^{-2}$, {\em (4)} sources with a strong FeK$\alpha$ line,
{\em (5)} sources located in the CT region of the $\rm L_x/L_{12}$ diagram, {\em (6)} sources located 
out of the $3\sigma$ region of the $\rm L_X/L_{[OIII]}$ diagram.\\
$\dagger$ The column density of the source is calculated by the Brightman \& Nandra (2011) model (see appendix A).} 
\end{minipage}
\end{table}

\subsubsection{The $\rm N_H$ vs. $L_x/L[OIII]$ ratio as an indicator of obscuration}

In this section we investigate the possibility that some of the sources are more obscured than we can infer from their column density.

\begin{figure}
\resizebox{9cm}{9cm}{\includegraphics{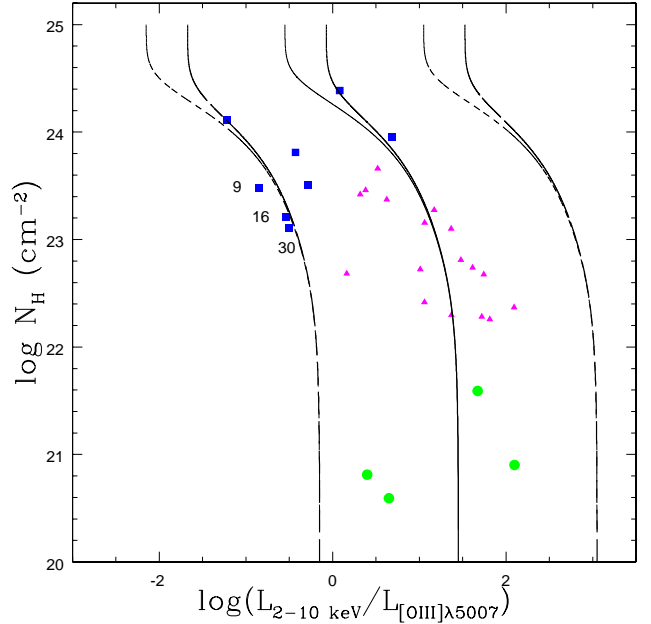}}
\caption{Distribution of the $\rm N_H$ values as a function of the
$L_{2-10 keV}/L_{[OIII]}$ ratio. Color- and shape-coding as in Fig. 1. The solid lines represent the mean
$\rm N_H$ vs. $L_{2-10 keV}/L_{[OIII]}$ relation followed by the Seyfert-1 sample in 
Akylas \& Georgantopoulos (2009) assuming a photon index of 1.8 and
3\% (thin line: 1\%) reflected radiation (see also Maiolino et al. 1998; Cappi
et al. 2006), while the dashed lines represent the $\pm3\sigma$ dispersion. 
The numbers of interesting sources correspond to the sequence numbers in Tables 1 and 2.}
\end{figure}

In Fig. 5 we plot the column density obtained from the
X-ray spectral modeling as a function of the X-ray to optical luminosity ratio. 
The [OIII] luminosities are corrected for reddening 
using the formula described in Basanni et al. (1999):
$\rm L_{[OIII]_{COR}} = L_{[OIII]_{OBS}} [(H\alpha /H\beta)/(H\alpha /H\beta)_o]^{2.94}$, where the intrinsic
Balmer decrement $\rm (H\alpha /H\beta )_o$ equals 3. The
lower left region in this plot could be possibly occupied by highly obscured 
or Compton-thick AGN, although their $\rm N_H$ values show the opposite (Akylas \& Georgantopoulos 2009).
In our case, however, none of the unobscured sources is located in this region, and therefore there 
is no evidence that their nuclei are heavily obscured. 

On the other hand, three sources with $\rm N_H>10^{23}$ cm$^{-2}$ are found 
marginally outside the 3$\sigma$ limit. This implies that they are probably even more 
obscured than what we calculated by fitting their X-ray spectra. Interestingly, these are the three
out of four sources (J103408.58+600152.1 (N9), J121839.40+470627.6 (N16), J215649.51--074532.4 (N30)) 
for which a high FeK$\alpha$ EW is reported, and they are also found
below the CT line in Fig. 4. Therefore, despite 
the value of the $\rm N_H$, it is evident that the iron line is a robust indicator of obscuration. 
Once again we can infer that our classification of unobscured and CT sources is valid.

\section{Discussion and conclusions}

\subsection{Candidate Compton-thick sources}

X-ray spectroscopy shows that the number of Compton-thick AGN in our sample could be as high as eight. 
N10 was initially included in the CT candidates because it is consistent with being CT within the uncertainties of the calculated column density.
However, we chose to exclude this source since it is not confirmed by any of the diagnostics presented in this study (see also LaMassa et al. 2014).
Therefore, we are left with seven CT sources, translating to a percentage of $\sim$23\%.

We find that the number of CT AGN found in our survey agrees with those in other X-ray surveys of optically selected Seyfert galaxies.
In more detail, Akylas \& Georgantopoulos (2009), using XMM-Newton observations, estimate the number of CT sources among the Seyfert galaxies from the 
Palomar spectroscopic sample of nearby galaxies (Ho, Filippenko \& Sargent 1995). They find a percentage of CT sources of 15-20 \%. 
Since their sample consists of nearby ($<$120 Mpc) Sy2 galaxies, the X-ray observations provide excellent spectra, hence accurate 
column density measurements classifications of all the AGN in their sample. Also, Malizia et al. (2009) reports that $\sim18\%$ of their 
hard X-ray selected Sy2 sample is Compton-thick. Nevertheless, considering only the low-redshift sources ($z<0.015$) 
to remove the selection bias that affects their sample against the detection of CT objects, the percentage becomes $\sim35\%$. 
They argue that this result is in excellent agreement with the percentage of CT AGN in the optically selected sample of Risaliti, Maiolino
\& Salvati (1999). 
We note that because of our sample selection, which requires a sufficient number of photons in order to derive X-ray 
spectra, we may also be biased against heavily obscured sources.

On the other hand, JJ13 in their SDSS optically selected sample of 
type 2 QSOs, estimate a higher percentage of CT sources that could be as high as 50\%, albeit with limited photon statistics.
Initially, the percentage they calculate based on the X-ray spectral modeling and the intensity of the FeK$\alpha$ line is significantly lower. However, 
it reaches 50\% after they conclude that at least half of the $\rm N_H$ values of their sources are underestimated, 
based on their $L_{2-10 keV}/L_{[OIII]}$ ratios. 

Nevertheless, four out of the seven CT sources in our study are in common with 
JJ13. Three of them are also reported as CT in JJ13. N12 is not a CT source in JJ13 despite its high $\rm N_H$ and the detection of the FeK$\alpha$ line
in their work. A probable reason is that they report an X-ray luminosity that is one order of
magnitude higher than the one we measure in the current study. Therefore the $L_{2-10 keV}/L_{[OIII]}$ ratio is higher than their threshold 
for a CT source.

We note that four sources in our sample were initially considered heavily obscured because of the high FeK$\alpha$ EW ($>$1 keV), 
although their column density was only a few times $10^{23}$ $\rm cm^{-2}$.
This suggests that these sources may be attenuated by CT absorbers. Indeed, all CT sources in the local Universe appear to present 
   high EW of the FeK$\alpha$ line (e.g., Fukazawa et al. 2011) owing to suppression of their continuum emission. 
   The discrepancy between the estimated column density and the EW  could be attributed to a more complex spectral 
   model that involves a double screen absorber with  one of them being 
   CT. In all four cases, by fitting the X-ray spectra
  with the model of Brightman \& Nandra (2011), we confirm that they are indeed 
  heavily obscured ($\rm N_H>10^{24} cm^{-2}$). In addition, according to Table 3, most of them satisfy all our CT criteria. 
      Interestingly, three out of the above four high-EW sources, J103408.58+600152.1 (N9), J215649.51--074532.4 (N30), and 
    J121839.40+470627.6 (N16) lie in the CT regime in the $\rm L_x/L_{12}$ diagram, and (Fig. 4) the same three sources 
    have the lowest $\rm L_X/L_{[OIII]}$ ratios (Fig. 5), again supporting their CT nature. Two of these sources
    are in common with JJ13 (N9 and N16), and present a high EW in both studies.

\subsection{Unabsorbed Sy2 nuclei}

The X-ray spectral analysis revealed that four\footnote{We note that (N25) is a flat X-ray spectrum source with a visible 
FeK$\alpha$ line, and we have excluded it from the list of unobscured sources (see appendix for more details)} 
Sy2 galaxies ($\sim$13\%) present very low absorption, below $10^{22}$ cm$^{-2}$, in sharp contrast
with the unification model of AGN. The percentage of unobscured Sy2 sources varies in the literature, from a few percent ($\sim 3-4\%$) 
in Risality, Maiolino \& Salvati (1999) and in Malizia et al. (2009), to 40\% in Page et al. (2006) and 66\% in Garcet et al. (2007). Our value 
is in better agreement with Panessa \& Bassani (2002) and Akylas \& Georgantopoulos (2009). However, considering that the number of unobscured Sy2s 
discovered in any of these studies is less than eight, we argue that we roughly agree with all of them, except perhaps with Garcet et al. (2007). 
Also, we note that our criteria for selecting narrow line AGN are more stringent than in most of the above studies; for example, 
Risality, Maiolino \& Salvati (1999) include Sy1.9 in their sample, and Garcet et al (2007) allow narrow line AGN up to FWHM$_{\ha}$=1500 km s$^{-1}$.

As we have already 
discussed, none of our unobscured sources present a low X-ray to
[OIII] or $\rm L_{12}$ luminosity ratio. They also do not present a strong FeK$\alpha$ line, and therefore we cannot 
associate them with a highly obscured Compton-thick nucleus.
In addition, the FWHM of their $\ha$ line is less than 500 km s$^{-1}$, which excludes the possibility 
of a narrow-line Sy1 classification. Although Tran (2001) argues about the presence of this kind of type 2 AGN in his sample of 
non-HBLR Sy2s fifteen years ago, their existence is still being strongly debated (see discussion in Antonucci 2012). Below, we summarize 
important observational and theoretical studies in the field, which attempt to approach this problem from various 
angles.

\begin{figure}
\resizebox{9cm}{9cm}{\includegraphics{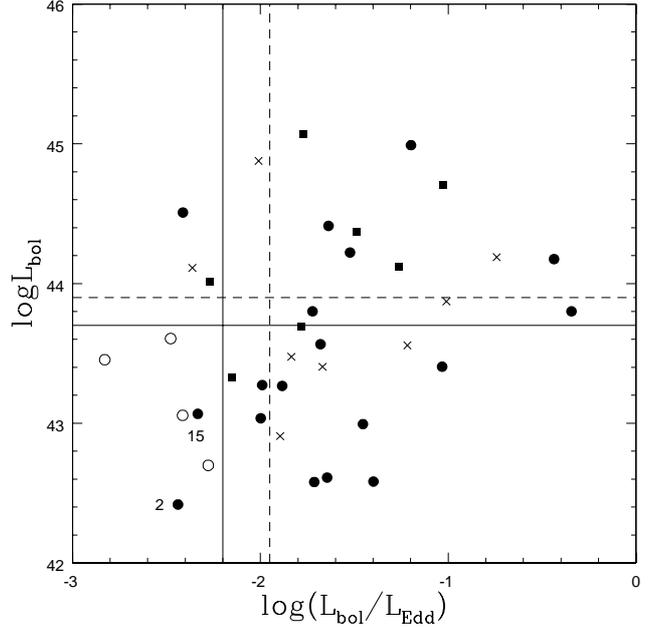}}
\caption{Eddington ratio vs. bolometric luminosity plot. The open circles denote the unabsorbed Sy2s, and the filled ones the absorbed ones. Squares denote 
the Compton-thick sources. The solid lines are the empirical thresholds found in the current work below which most of our unobscured Sy2s are found. 
The dashed lines are the empirical thresholds shown in Marinucci et al. (2012) that separate HBLR and non-HBLR Sy2s.}
\end{figure}

There is strong evidence that the dusty obscuring torus in low luminosity AGN is
absent or is thinner than expected in higher luminosities (e.g., Elitzur \& Shlosman 2006; Perlman et al. 2007; van der Wolk et
al. 2010). Accordingly, all low luminosity AGN should have been Type 1 sources,
which of course is not the case. The only reasonable explanation of this problem
is the additional absence of the BLR in such systems. 
Some authors (e.g., Nicastro 2000; Nicastro, Martocchia \& Matt 2003; Bian \& Gu. 2007; Marinucci et al. 2012; Elitzur, Ho \& Trump 2014)
presented arguments that below a specific accretion rate of material into the
black hole, and therefore at lower luminosities, the BLR might also be absent.

Using data from nearby bright AGN, Elitzur \& Ho (2009) conclude that the BLR
disappears at bolometric luminosities that are lower than $5 \times 10^{39} (M_{BH}/10^7 M_{\sun})^{2/3}
\rm erg\; s^{-1}$, where $M_{BH}$ is the mass of the black hole. They also argue
that the quenching of the BLR and the disappearance of the torus can occur
either simultaneously or in sequence, with decreasing black hole accretion rate and
luminosity. Thus, a possible scenario would be that non-HBLR Sy2 AGN are objects 
lacking the BLR and possibly the torus. Nicastro, Martocchia \& Matt (2003) conclude  
that the BLR probably does not exist below an accretion rate threshold of $\log(L_{bol}/L_{Edd})=-3$,
while Marinucci et al. (2012) argue that true Sy2s can be found below the relatively higher limits of 
bolometric luminosity $\log L_{bol}=43.9$ and Eddington ratio $\log(L_{bol}/L_{Edd})=-1.9$. 
Marinucci et al. (2012) derived the bolometric luminosity from the X-ray and the 
[OIV] luminosity and conclude that $L_{[\rm OIII]}$ is not as reliable (see also relevant discussion in Elitzur 2012). 
We note that Elitzur \& Ho (2009) thresholds are relatively low, not only compared to other studies but also for the general Sy2 population (see 
discussion in the recent review by Netzer 2015). However, the idea that the accretion rate is essential in the formation of the BLR 
seems to be valid, although the exact limits have not yet been defined and probably also depend on other factors (see discussion in Koulouridis 2014).  

 To evaluate the above limits for our four unobscured sources, we computed their bolometric luminosities from the SED modeling (see \S4.2.4). 
We also calculated their black hole masses using the $M_{BH}-\sigma*$ relation (Tremaine et al. 2002), where $\sigma*$ is the stellar velocity dispersion, 
calculated from the FWHM of the [OIII] emission lines (Greene \& Ho 2005). We find that the Elitzur \& Ho (2009) limits are
very low for our unobscured sources. Nevertheless, all satisfy the bolometric luminosity and Eddington ratio limit 
of Marrinucci et al. (2012). We note, however, that our Eddington ratios may be overestimated since the Eddington luminosities, 
derived from the FWHM of the [OIII] lines, are probably underestimated (e.g., Bian \& Gu 2007).

By conducting a two-sample Student's t-test
between the accretion rates of the unobscured and the obscured sources, we conclude that their mean values are significantly different at the 99.9\% 
confidence level. In Fig. 6 we plot the 
Eddington ratio versus the bolometric luminosity of our Sy2s, but also the discarded intermediate type Seyferts (crosses). 
We also plot the lines that apparently separate the unobscured sources from the rest of the Sy2 
population. These limits are similar to the respective ones found by Marinucci et al. (dashed lines in Fig.6) 
for HBLR and non-HBLR sources. All four unobscured sources fall into the area where non-HBLR Sy2s are found and the BLR is predicted to not exist. We note 
that the limits of previous works were based on the differences between HBLR and non-HBLR Sy2s, while our sample is divided into obscured and unobscured 
sources. The unobscured Sy2s are non-HBLR Sy2s by definition, whereas the obscured sources are not necessarily HBLR Sy2s. 
Therefore, the presence of obscured Sy2s in the bottom left quarter of the plot may imply the lack of their BLR as well.
Interestingly, a number of Compton-thick sources exhibit low accretion rates. This agrees with the evolutionary scheme 
of AGN proposed by Koulouridis (2014), where a fraction of Compton-thick 
sources are predicted to emerge shortly after a galaxy interaction or merging event that causes the inflow of gas and dust toward 
the central region of the galaxy,
enhances circumnuclear star formation and triggers the AGN. During this phase the accretion rate is expected to be low and the BLR absent. 
However, the failure to detect the BLR in CT sources may as well be due to the heavy obscuration and the large covering factor 
of the nucleus (see next paragraph).  We note that the uncertainties that enter the above calculations are large (see Greene \& Ho 2005) and 
our samples fairly small. However, the general tendency of low accretion type-2 AGN to lack 
any evidence of a BLR is once more evident.

An alternative scenario that can explain the lack of detectable BLR in many CT sources is that heavy obscuration does not 
allow the detection of the BLR even in the polarized spectrum.
Marinucci et al. (2012) conclude that 64\% of
their compton-thick non-HBLR Sy2s exhibit higher accretion rates than the
threshold clearly separating the two Sy2 classes. They attributed this
discrepancy to heavy absorption along our line of sight, preventing the
detection of the actual BLR in their nuclei. Evidently, merging systems constitute
a class of extragalactic objects where heavy obscuration occurs (e.g., Hopkins et
al. 2008). The merging process may also lead to rapid black hole growth, giving
birth to a heavily absorbed and possibly Compton-thick AGN. Thus, we could
presume that a number of our non-HBLR mergers, if not all of them, might actually be
BLR AGN galaxies, where the high concentration of gas and dust prohibits even
the indirect detection of the broad line emission (e.g., Shu et al. 2007). However, other studies 
have concluded that there is no evidence that non-HBLR Sy2s are more obscured than their HBLR peers (Tran 2003; Yu 2005; Wu 2011), while 
totally unobscured low-luminosity non-HBLR Sy2s were detected via investigation 
of their X-ray properties (e.g., Panessa \& Bassani 2002; Akylas \& Georgantopoulos 2009). The total population of non-HBLR Sy2s
is probably a mixture of objects with low accretion rate and/or high obscuration. 

Koulouridis (2014) argue that both of the above scenarios agree with an AGN evolutionary scheme
(Krongold et al. 2002; Koulouridis et al. 2006a, b, 2013), 
where a low accretion rate is predicted at the beginning and the end of the Seyfert 
duty cycle, without ruling out the possibility that some HBLR Sy2s
could also be created by minor disturbances or even secular processes.

Finally, we note that there is always the possibility that the discrepancy 
between the optical and the X-ray spectra is due to variability, since they were not obtained 
simultaneously.\\
 
In a nutshell: 
\begin{enumerate}
 \item We found four unobscured sources ($\sim$13\%) at odds with the simplest unification scheme. These 
 sources exhibit low accretion rates that agree with previous studies that predict the lack of the BLR in low-accretion-rate AGN.
 \item 64\% of the Sy2s are obscured with a median column density value of $\rm N_H\sim1.0\times10^{23}cm^{-2}$. 
 \item The percentage of CT AGN is at $\sim$23\%,
although direct comparison with previous studies is difficult because of the different selection methodologies. 
Their heavy obscuration was confirmed using a variety of criteria and diagnostics.  
\end{enumerate}

\acknowledgements
We thank the anonymous referee for the insightful comments and suggestions that significantly contributed to improving the quality of the publication.
EK acknowledges fellowship funding provided by the Greek General Secretariat of Research
and Technology in the framework of the program Support of Postdoctoral
Researchers, PE-1145.
This work is based on observations obtained with XMM-Newton, an ESA science
mission with instruments and contributions directly funded by ESA Member States
and the USA (NASA). Funding for SDSS-III has been provided by the Alfred P. Sloan Foundation, the
Participating Institutions, the National Science Foundation, and the U.S.
Department of Energy Office of Science. The SDSS-III web site is
http://www.sdss3.org/.

\appendix
\section{notes on individual objects}

\begin{itemize}

\item  Source 9 - J103408.58+600152.1\newline 
 Because of the large EW of the FeK$\alpha$ line, but the relatively low $\rm N_H$, we fit the spectrum with the model of 
 Brightman \& Nandra (2011).
 The result of the fit is a high column density, $\rm N_H=220^{+\infty}_{-70}$, characteristic of the CT 
 sources. Other useful values: p1/p2=0.008, cstat/dof=51.9/34, $\Gamma_{soft}=3^{+0.2}_{-0.4}$,
$\Gamma_{hard}=1.8$ (fixed).
\\
\item  Source 16 - J121839.40+470627.6\newline 
 Because of the large EW of the FeK$\alpha$ line, but the relatively low $\rm N_H$, we fit the spectrum with the model of 
 Brightman \& Nandra (2011).
 The result of the fit is a high column density, $\rm N_H=2009^{+\infty}_{-135}$, characteristic of the CT 
 sources. Other useful values: p1/p2=0.003, cstat/dof=145/155, $\Gamma_{soft}=3.4^{+0.8}_{-0.6}$,
$\Gamma_{hard}=1.8$ (fixed).
\\
\item Source 22 - J131104.66+272807.2 \newline 
 Because of the high $\rm N_H$, but the small EW of the FeK$\alpha$ line, we fit the spectrum with the model of 
 Brightman \& Nandra (2011).
 The result of the fit is a column density value of $\rm N_H=114^{+87}_{-29}\times10^{22}$, which is relatively lower than what is reported 
 in the current study, but again above the limit that characterize CT sources. 
 A strong FeK$\alpha$ line is only present in the pn detector.
 Other useful values: p1/p2=0.003, cstat/dof=300/366, $\Gamma_{soft}=2.6^{+0.2}_{-0.2}$,
$\Gamma_{hard}=1.8$ (fixed).
\\
\item Source 25 - J135436.29+051524.5\newline
We chose not to include this source in the unabsorbed list because its
 photon index $\Gamma$ is extremely flat ($\sim$0.8) if left as a free parameter, 
 and in addition there seems to be a strong FeK$\alpha$ line. It may be a reflection
-dominated Compton-thick source, but we cannot confirm this because of the relatively low quality X-ray spectrum. 
 Also, even though the EW seems high, it cannot be considered as 
 a Compton-thick candidate because the scattered percentage is too large ($>30$\%)
 implying partial covering instead of scattered emission.
 \\
 \item Source 29 - J150754.38+010816.8\newline 
 Because of the large EW of the FeK$\alpha$ line, but the relatively low $\rm N_H$, we fit the spectrum with the model of 
 Brightman \& Nandra (2011).
 The result of the fit is a high column density, $\rm N_H=211^{+\infty}_{-61}$, characteristic of the CT 
 sources. Other useful values: p1/p2=0.003, cstat/dof=320/355, $\Gamma_{soft}=3.2^{+0.4}_{-0.4}$,
$\Gamma_{hard}=1.8$ (fixed).
\\
\item Source 30 - J215649.51--074532.4\newline 
 Because of the large EW of the FeK$\alpha$ line, but the relatively low $\rm N_H$, we fit the spectrum with the model of 
 Brightman \& Nandra (2011).
 The result of the fit is a high column density, $\rm N_H=1500^{+\infty}_{-1200}$, which is characteristic of the CT 
 sources. Other useful values: p1/p2=0.003, cstat/dof=265/3, $\Gamma_{soft}=3.5^{+0.5}_{-0.5}$,
$\Gamma_{hard}=1.8$ (fixed).
\end{itemize}


\begin{thebibliography} {}
\bibitem[Akylas 
\& Georgantopoulos(2009)]{2009A&A...500..999A} Akylas, A., \& Georgantopoulos, I.\ 2009, \aap, 500, 999 
\bibitem[Akylas et 
al.(2012)]{2012A&A...546A..98A} Akylas, A., Georgakakis, A., Georgantopoulos, I., Brightman, M., \& Nandra, K.\ 2012, \aap, 546, A98 
\bibitem[Antonucci(1993)]{1993ARA&A..31..473A} Antonucci, R.\ 1993, ARA\&A, 31,
473 
\bibitem[Antonucci(2012)]{2012A&AT...27..557A} Antonucci, R.\ 2012, Astronomical and Astrophysical Transactions, 27, 557 
\bibitem[Arnaud(1996)]{1996ASPC..101...17A} Arnaud, K.~A.\ 1996, 
Astronomical Data Analysis Software and Systems V, 101, 17 
\bibitem[Asari et al.(2007)]{2007MNRAS.381..263A} Asari, N.~V., Cid 
Fernandes, R., Stasi{\'n}ska, G., et al.\ 2007, \mnras, 381, 263 
\bibitem[Awaki et al.(2000)]{2000ApJ...542..175A} Awaki, H., Ueno, S., 
Taniguchi, Y., \& Weaver, K.~A.\ 2000, \apj, 542, 175 
\bibitem[Baldwin et al.(1981)]{1981PASP...93....5B} Baldwin, J.~A., 
Phillips, M.~M., \& Terlevich, R.\ 1981, \pasp, 93, 5 
\bibitem[Bassani et al.(1999)]{1999ApJS..121..473B} Bassani, L., Dadina, 
M., Maiolino, R., et al.\ 1999, \apjs, 121, 473
\bibitem[Bian 
\& Gu(2007)]{2007ApJ...657..159B} Bian, W., \& Gu, Q.\ 2007, \apj, 657, 159 
\bibitem[Brandt 
\& Alexander(2015)]{2015A&ARv..23....1B} Brandt, W.~N., \& Alexander, D.~M.\ 2015, \aapr, 23, 1 
\bibitem[Brinchmann et al.(2004)]{2004astro.ph..6220B} Brinchmann, J., 
Charlot, S., Heckman, T.~M., et al.\ 2004, arXiv:astro-ph/0406220 
\bibitem[Brightman 
\& Nandra(2011)]{2011MNRAS.414.3084B} Brightman, M., \& Nandra, K.\ 2011, \mnras, 414, 3084
\bibitem[Bruzual 
\& Charlot(2003)]{2003MNRAS.344.1000B} Bruzual, G., \& Charlot, S.\ 2003, \mnras, 344, 1000 
\bibitem[Caccianiga et 
al.(2007)]{2007A&A...470..557C} Caccianiga, A., Severgnini, P., Della Ceca, R., et al.\ 2007, \aap, 470, 557
\bibitem[Cappi et 
al.(2006)]{2006A&A...446..459C} Cappi, M., Panessa, F., Bassani, L., et al.\ 2006, \aap, 446, 459 
\bibitem[Cardelli et al.(1989)]{1989ApJ...345..245C} Cardelli, J.~A., 
Clayton, G.~C., \& Mathis, J.~S.\ 1989, \apj, 345, 245 
\bibitem[Churazov et 
al.(2007)]{2007A&A...467..529C} Churazov, E., Sunyaev, R., Revnivtsev, M., et al.\ 2007, \aap, 467, 529 
\bibitem[Cid Fernandes et al.(2005)]{2005MNRAS.358..363C} Cid Fernandes, 
R., Mateus, A., Sodr{\'e}, L., Stasi{\'n}ska, G., 
\& Gomes, J.~M.\ 2005, \mnras, 358, 363 
\bibitem[Cid Fernandes et al.(2007)]{2007MNRAS.375L..16C} Cid Fernandes, 
R., Asari, N.~V., Sodr{\'e}, L., et al.\ 2007, \mnras, 375, L16 
\bibitem[Cid Fernandes et al.(2011)]{2011MNRAS.413.1687C} Cid Fernandes, 
R., Stasi{\'n}ska, G., Mateus, A., 
\& Vale Asari, N.\ 2011, \mnras, 413, 1687 
\bibitem[Comastri(2004)]{2004ASSL..308..245C} Comastri, A.\ 2004, 
Supermassive Black Holes in the Distant Universe, 308, 245 
\bibitem[Corral et 
al.(2015)]{2015A&A...576A..61C} Corral, A., Georgantopoulos, I., Watson, M.~G., et al.\ 2015, \aap, 576, A61 
\bibitem[Dickey 
\& Lockman(1990)]{1990ARA&A..28..215D} Dickey, J.~M., \& Lockman, F.~J.\ 1990, \araa, 28, 215 
\bibitem[Dultzin-Hacyan et al.(1999)]{1999ApJ...513L.111D}
  Dultzin-Hacyan, D., et al. \ 1999, \apjl, 513, L111 
\bibitem[Elitzur 
\& Shlosman(2006)]{2006ApJ...648L.101E} Elitzur, M., \& Shlosman, I.\ 2006,
\apjl, 648, L101 
\bibitem[Elitzur 
\& Ho(2009)]{2009ApJ...701L..91E} Elitzur, M., \& Ho, L.~C.\ 2009, \apjl, 701, L91 
\bibitem[Elitzur et al.(2014)]{2014MNRAS.438.3340E} Elitzur, M., Ho, L.~C., 
\& Trump, J.~R.\ 2014, \mnras, 438, 3340 
\bibitem[Frontera et al.(2007)]{2007ApJ...666...86F} Frontera, F., 
Orlandini, M., Landi, R., et al.\ 2007, \apj, 666, 86 
\bibitem[Fukazawa et al.(2011)]{2011ApJ...727...19F} Fukazawa, Y., Hiragi, 
K., Mizuno, M., et al.\ 2011, \apj, 727, 19 
\bibitem[Garcet et 
al.(2007)]{2007A&A...474..473G} Garcet, O., Gandhi, P., Gosset, E., et al.\ 2007, \aap, 474, 473 
\bibitem[Georgakakis 
\& Nandra(2011)]{2011MNRAS.414..992G} Georgakakis, A., \& Nandra, K.\ 2011, \mnras, 414, 992 
\bibitem[Georgantopoulos et 
al.(2013)]{2013A&A...555A..43G} Georgantopoulos, I., Comastri, A., Vignali, C., et al.\ 2013, \aap, 555, AA43
\bibitem[Georgantopoulos(2013)]{2013IJMPS..23....1G} Georgantopoulos, I.\ 
2013, International Journal of Modern Physics Conference Series, 23, 1 
\bibitem[Gandhi et 
al.(2009)]{2009A&A...502..457G} Gandhi, P., Horst, H., Smette, A., et al.\ 2009, \aap, 502, 457 
\bibitem[Ghisellini et al.(1991)]{1991MNRAS.248...14G} Ghisellini, G., 
George, I.~M., Fabian, A.~C., \& Done, C.\ 1991, \mnras, 248, 14 
\bibitem[Gilli et 
al.(2007)]{2007A&A...463...79G} Gilli, R., Comastri, A., \& Hasinger, G.\ 2007, \aap, 463, 79 
\bibitem[Goulding et al.(2011)]{2011MNRAS.411.1231G} Goulding, A.~D., 
Alexander, D.~M., Mullaney, J.~R., et al.\ 2011, \mnras, 411, 1231 
\bibitem[Greene 
\& Ho(2005)]{2005ApJ...627..721G} Greene, J.~E., \& Ho, L.~C.\ 2005, \apj, 627, 721 
\bibitem[Heckman et al.(2005)]{2005ApJ...634..161H} Heckman, T.~M., Ptak, 
A., Hornschemeier, A., \& Kauffmann, G.\ 2005, \apj, 634, 161 
\bibitem[Ho et al.(1995)]{1995ApJS...98..477H} Ho, L.~C., Filippenko, 
A.~V., \& Sargent, W.~L.\ 1995, \apjs, 98, 477 
\bibitem[Ho(2008)]{2008ARA&A..46..475H} Ho, L.~C.\ 2008, \araa, 46, 475 
\bibitem[Hopkins et al.(2008)]{2008ApJS..175..356H} Hopkins, P.~F., 
Hernquist, L., Cox, T.~J., \& Kere{\v s}, D.\ 2008, \apjs, 175, 356
\bibitem[Hunt 
\& Malkan(1999)]{1999ApJ...516..660H} Hunt, L.~K., \& Malkan, M.~A.\ 1999, \apj,
516, 660 
\bibitem[Jia et al.(2013)]{2013ApJ...777...27J} Jia, J., Ptak, A., Heckman, 
T., \& Zakamska, N.~L.\ 2013, \apj, 777, 27 \
\bibitem[Kauffmann et al.(2003)]{2003MNRAS.346.1055K} Kauffmann, G., 
Heckman, T.~M., Tremonti, C., et al.\ 2003, \mnras, 346, 1055 
\bibitem[Kewley et al.(2001)]{2001ApJ...556..121K} Kewley, L.~J., Dopita, 
M.~A., Sutherland, R.~S., Heisler, C.~A., 
\& Trevena, J.\ 2001, \apj, 556, 121 
\bibitem[Koulouridis et al.(2006)]{2006ApJ...639...37K} Koulouridis,
  E., et al. \ 2006a, \apj, 639, 37 
\bibitem[Koulouridis et al.(2006)]{2006ApJ...651...93K} Koulouridis,
  E., et al. \ 2006b, \apj, 651, 93 
\bibitem[Koulouridis et 
al.(2013)]{2013A&A...552A.135K} Koulouridis, E., Plionis, M., Chavushyan, V., et al.\ 2013, \aap, 552, A135 
\bibitem[Koulouridis(2014)]{2014A&A...570A..72K} Koulouridis, E.\ 2014, \aap, 570, AA72 
\bibitem[Krongold et al.(2002)]{2002ApJ...572..169K} Krongold, Y., 
Dultzin-Hacyan, D., \& Marziani, P.\ 2002, \apj, 572, 169 
\bibitem[Krumpe et 
al.(2008)]{2008A&A...483..415K} Krumpe, M., Lamer, G., Corral, A., et al.\ 2008, \aap, 483, 415 
\bibitem[LaMassa et al.(2009)]{2009ApJ...705..568L} LaMassa, S.~M., 
Heckman, T.~M., Ptak, A., et al.\ 2009, \apj, 705, 568 
\bibitem[LaMassa et al.(2014)]{2014ApJ...787...61L} LaMassa, S.~M., Yaqoob, 
T., Ptak, A.~F., et al.\ 2014, \apj, 787, 61 
\bibitem[Lamastra et 
al.(2009)]{2009A&A...504...73L} Lamastra, A., Bianchi, S., Matt, G., et al.\ 2009, \aap, 504, 73 
\bibitem[Leahy 
\& Creighton(1993)]{1993MNRAS.263..314L} Leahy, D.~A., \& Creighton, J.\ 1993, \mnras, 263, 314 
\bibitem[Levenson et al.(2001)]{2001ApJ...550..230L} Levenson, N.~A., 
Weaver, K.~A., \& Heckman, T.~M.\ 2001, \apj, 550, 230
\bibitem[Lutz et 
al.(2004)]{2004A&A...418..465L} Lutz, D., Maiolino, R., Spoon, H.~W.~W., \& Moorwood, A.~F.~M.\ 2004, \aap, 418, 465 
\bibitem[Maiolino et 
al.(1998)]{1998A&A...338..781M} Maiolino, R., Salvati, M., Bassani, L., et al.\ 1998, \aap, 338, 781 
\bibitem[Maiolino et 
al.(2007)]{2007A&A...468..979M} Maiolino, R., Shemmer, O., Imanishi, M., et al.\ 2007, \aap, 468, 979 
\bibitem[Malizia et al.(2009)]{2009MNRAS.399..944M} Malizia, A., Stephen, 
J.~B., Bassani, L., et al.\ 2009, \mnras, 399, 944 
\bibitem[Marconi et al.(2004)]{2004MNRAS.351..169M} Marconi, A., Risaliti, 
G., Gilli, R., et al.\ 2004, \mnras, 351, 169
\bibitem[Marinucci et al.(2012)]{2012ApJ...748..130M} Marinucci, A., 
Bianchi, S., Nicastro, F., Matt, G., 
\& Goulding, A.~D.\ 2012, \apj, 748, 130 
\bibitem[Mateus et al.(2006)]{2006MNRAS.370..721M} Mateus, A., Sodr{\'e}, 
L., Cid Fernandes, R., et al.\ 2006, \mnras, 370, 721
\bibitem[Matt et al.(2000)]{2000MNRAS.318..173M} Matt, G., Fabian, A.~C., 
Guainazzi, M., et al.\ 2000, \mnras, 318, 173 
\bibitem[Miller 
\& Antonucci(1983)]{1983ApJ...271L...7M} Miller, J.~S., \& Antonucci, R.~R.~J.\ 1983, \apjl, 271, L7
\bibitem[Mulchaey et al.(1994)]{1994ApJ...436..586M} Mulchaey, J.~S., 
Koratkar, A., Ward, M.~J., et al.\ 1994, \apj, 436, 586 
\bibitem[Netzer(2015)]{2015arXiv150500811N} Netzer, H.\ 2015, 
arXiv:1505.00811 
\bibitem[Nicastro(2000)]{2000ApJ...530L..65N} Nicastro, F.\ 2000, \apjl, 
530, L65
\bibitem[Nicastro et al.(2003)]{2003ApJ...589L..13N} Nicastro, F., 
Martocchia, A., \& Matt, G.\ 2003, \apjl, 589, L13
\bibitem[Panessa
\& Bassani(2002)]{2002A&A...394..435P} Panessa, F., \& Bassani, L.\ 2002, \aap,
394, 435 
\bibitem[Panessa et 
al.(2006)]{2006A&A...455..173P} Panessa, F., Bassani, L., Cappi, M., et al.\ 2006, \aap, 455, 173 
\bibitem[Page et al.(2006)]{2006MNRAS.369..156P} Page, M.~J., Loaring, 
N.~S., Dwelly, T., et al.\ 2006, \mnras, 369, 156 
\bibitem[Pappa et al.(2000)]{2000MNRAS.314..589P} Pappa, A., 
Georgantopoulos, I., \& Stewart, G.~C.\ 2000, \mnras, 314, 589
\bibitem[Perlman et al.(2007)]{2007ApJ...663..808P} Perlman, E.~S., Mason, 
R.~E., Packham, C., et al.\ 2007, \apj, 663, 808 
\bibitem[Ptak et al.(2006)]{2006ApJ...637..147P} Ptak, A., Zakamska, N.~L., 
Strauss, M.~A., et al.\ 2006, \apj, 637, 147 
\bibitem[Risaliti et al.(1999)]{1999ApJ...522..157R} Risaliti, G., 
Maiolino, R., \& Salvati, M.\ 1999, \apj, 522, 157 
\bibitem[Risaliti(2002)]{2002A&A...386..379R} Risaliti, G.\ 2002, \aap, 386, 379 
\bibitem[Rosen et al.(2015)]{2015arXiv150407051R} Rosen, S.~R., Webb, 
N.~A., Watson, M.~G., et al.\ 2015, arXiv:1504.07051 
\bibitem[Rovilos et al.(2014)]{2014MNRAS.438..494R} Rovilos, E., 
Georgantopoulos, I., Akylas, A., et al.\ 2014, \mnras, 438, 494
\bibitem[Schawinski et al.(2007)]{2007MNRAS.382.1415S} Schawinski, K., 
Thomas, D., Sarzi, M., et al.\ 2007, \mnras, 382, 1415 
\bibitem[Schlegel et al.(1998)]{1998ApJ...500..525S} Schlegel, D.~J., 
Finkbeiner, D.~P., \& Davis, M.\ 1998, \apj, 500, 525 
\bibitem[Shu et al.(2007)]{2007ApJ...657..167S} Shu, X.~W., Wang, J.~X., 
Jiang, P., Fan, L.~L., \& Wang, T.~G.\ 2007, \apj, 657, 167 
\bibitem[Smith 
\& Done(1996)]{1996MNRAS.280..355S} Smith, D.~A., \& Done, C.\ 1996, \mnras, 280, 355 
\bibitem[Str{\"u}der et 
al.(2001)]{2001A&A...365L..18S} Str{\"u}der, L., Briel, U., Dennerl, K., et al.\ 2001, \aap, 365, L18 
\bibitem[Thomas et al.(2013)]{2013MNRAS.431.1383T} Thomas, D., Steele, O., 
Maraston, C., et al.\ 2013, \mnras, 431, 1383 
\bibitem[Tran(2001)]{2001ApJ...554L..19T} Tran, H.~D.\ 2001, \apjl, 554, 
L19
\bibitem[Tran(2003)]{2003ApJ...583..632T} Tran, H.~D.\ 2003, \apj,
  583, 632
\bibitem[Treister et al.(2009)]{2009ApJ...696..110T} Treister, E., Urry, 
C.~M., \& Virani, S.\ 2009, \apj, 696, 110 
\bibitem[Tremaine et al.(2002)]{2002ApJ...574..740T} Tremaine, S., 
Gebhardt, K., Bender, R., et al.\ 2002, \apj, 574, 740 
\bibitem[Tremonti et al.(2004)]{2004ApJ...613..898T} Tremonti, C.~A., 
Heckman, T.~M., Kauffmann, G., et al.\ 2004, \apj, 613, 898 
\bibitem[Trouille 
\& Barger(2010)]{2010ApJ...722..212T} Trouille, L., \& Barger, A.~J.\ 2010, \apj, 722, 212 
\bibitem[Turner et al.(1997)]{1997ApJS..113...23T} Turner, T.~J., George, 
I.~M., Nandra, K., \& Mushotzky, R.~F.\ 1997, \apjs, 113, 23 
\bibitem[Turner et 
al.(2001)]{2001A&A...365L..27T} Turner, M.~J.~L., Abbey, A., Arnaud, M., et al.\ 2001, \aap, 365, L27 
\bibitem[van der Wolk et 
al.(2010)]{2010A&A...511A..64V} van der Wolk, G., Barthel, P.~D., Peletier,
R.~F., \& Pel, J.~W.\ 2010, \aap, 511, A64
\bibitem[V{\'e}ron-Cetty 
\& V{\'e}ron(2010)]{2010A&A...518A..10V} V{\'e}ron-Cetty, M.-P., \& V{\'e}ron, P.\ 2010, \aap, 518, A10 
\bibitem[Villarroel 
\& Korn(2014)]{2014NatPh..10..417V} Villarroel, B., \& Korn, A.~J.\ 2014, Nature Physics, 10, 417 
\bibitem[Watson et 
al.(2009)]{2009A&A...493..339W} Watson, M.~G., Schr{\"o}der, A.~C., Fyfe, D., et al.\ 2009, \aap, 493, 339 
\bibitem[Weaver et al.(1996)]{1996ApJ...458..160W} Weaver, K.~A., Nousek, 
J., Yaqoob, T., et al.\ 1996, \apj, 458, 160 
\bibitem[Wright et al.(2010)]{2010AJ....140.1868W} Wright, E.~L., 
Eisenhardt, P.~R.~M., Mainzer, A.~K., et al.\ 2010, \aj, 140, 1868 
\bibitem[Wu et al.(2011)]{2011ApJ...730..121W} Wu, Y.-Z., Zhang, E.-P., 
Liang, Y.-C., Zhang, C.-M., \& Zhao, Y.-H.\ 2011, \apj, 730, 121 
\bibitem[Yu 
\& Hwang(2005)]{2005ApJ...631..720Y} Yu, P.-C., \& Hwang, C.-Y.\ 2005, \apj, 631, 720 
\end{thebibliography}
\end{document}